\documentclass[fleqn,10pt]{wlscirep}
\usepackage[utf8]{inputenc}
\usepackage[T1]{fontenc}
\title{Failure of neuron network coherence induced by SARS-CoV-2-infected astrocytes}

\author[1,*]{Sergey V. Stasenko}
\author[2,3]{Alexander E. Hramov}
\author[1,2,4]{Victor B. Kazantsev}
\affil[1]{Laboratory of advanced methods for high-dimensional data analysis, Lobachevsky university,  Nizhniy Novgorod,603022, Russia}
\affil[2]{Neuroscience and Cognitive Technology Lab, Innopolis University, Innopolis, 420500, Russia}
\affil[3]{Department of New Cardiological Informational Technologies, Saratov State Medical University, Saratov, 410012, Russia}
\affil[4]{Laboratory for Neuromodeling, Neuroscience Research Institute, Samara State Medical University, Samara, 443099, Russia}
\affil[*]{stasenko@neuro.nnov.ru}


\begin{abstract}
Coherent activations of brain neuron networks underlay many physiological functions associated with various behavioral states. These synchronous fluctuations in the electrical activity of the brain are also referred to as brain rhythms. At the cellular level, the rhythmicity can be induced by various mechanisms of intrinsic oscillations in neurons or network circulation of excitation between synaptically coupled neurons. One of the specific mechanisms concerns the activity of brain astrocytes that accompany neurons and can coherently modulate synaptic contacts of neighboring neurons, synchronizing their activity. Recent studies have shown that coronavirus infection (Covid-19), entering the central nervous system and infecting astrocytes, causes various metabolic disorders. Specifically, Covid-19 can depress the synthesis of astrocytic glutamate and GABA. It is also known that in the postcovid state, patients may suffer from symptoms of anxiety and impaired cognitive functions, which may be a consequence of disturbed brain rhythms. We propose a mathematical model of a spiking neural network accompanied by astrocytes capable to generate quasi-synchronous rhythmic bursting discharges. The model predicts that if the astrocytes are infected, and the release of glutamate is depressed, then normal burst rhythmicity suffers dramatically. Interestingly, in some cases, the failure of network coherence may be intermittent with intervals of normal rhythmicity, or the synchronization can completely disappears.
\end{abstract}
\begin{document}

\flushbottom
\maketitle
%
%
\thispagestyle{empty}


\section*{Introduction}

The synchronization of the neural network activity at the cellular and network levels gives rise to rhythmic voltage fluctuations traveling across brain regions, known as neuronal oscillations or brain waves \cite{Buzsaki2006,Buskila2019}. Modulation of neural oscillations is provided by the dynamic interplay between neuronal connectivity patterns, cellular membrane properties, intrinsic circuitry, speed of axonal conduction, and synaptic delays \cite{Nunez1996,Cunningham2006,Buskila2013, Tapson2013}. The neural oscillations fluctuate between two main states, known as “up states” and “down states” \cite{Sanchez-Vives2000}. The network coherence providing by the up state in spatially organized cortical neural ensembles play a crucial role for several sensory and motor processes, as well as for cognitive flexibility (i.e., attention, memory), thereby playing a fundamental role in the brain’s basic functions \cite{Fries2001,Tallon-Baudry2004}. Furthermore, different network dynamics (from slow to ultra-fast oscillations) can change according to the behavioral state, with some frequency bands being associated with sleep, while other frequencies predominate during arousal or conscious states \cite{Brooks1968, Achermann1997,Murthy1992}.

Besides purely neuronal mechanisms, many recent studies revealed the essential contributions made by astrocytes to many physiological brain functions,including synaptogenesis \cite{Ullian2001}, metabolic coupling \cite{Magistretti2006}, nitrosative regulation of synaptic release \cite{Buskila2005, Abu-Ghanem2008,Buskila2010}, synaptic transmission \cite{Fields2002}, network oscillations \cite{Bellot-Saez2018}, and plasticity \cite{Suzuki2011, Oberheim2012}. Astrocytes can play a significant role in the processing of synaptic information through impact on pre- and post-synaptic neurons. This fact leads to the concept of a tripartite synapse \cite{Araque1999,Haydon2001}. A part of the neurotransmitter released from the presynaptic terminals (i.e., glutamate) can diffuse out of the synaptic cleft and bind to metabotropic glutamate receptors (mGluRs) on the astrocytic processes that are located near the neuronal synaptic compartments. The neurotransmitter activates G-protein mediated signaling cascades that result in phospholipase C (PLC) activation and insitol-1,4,5-trisphosphaste (IP3) production. The IP3 binds to IP3-receptors in the intracellular stores and triggers $Ca^{2+}$ release into the cytoplasm. Such an increase in intracellular $Ca^{2+}$ can trigger the release of gliotransmitters \cite{Parpura2010} [e.g., glutamate, adenosine triphosphate (ATP), D-serine, and GABA] into the extracellular space. A gliotransmitter can affect both the pre- and post-synaptic parts of the neuron. By binding to presynaptic receptors, it can either potentiate or depress presynaptic release probability. One of the key pathways in the tripartite synapse is mediated by glutamate released by the astrocyte \cite{Parri2001, Liu2004, Perea2007}. Such glutamate can potentially target presynaptic NMDA receptors, which increase the release probability \cite{McGuinness2010}, or presynaptic mGluRs, which decrease it \cite{Semyanov2000}. Presynaptic kainate receptors exhibit a more complex modulation of synaptic transmission through both metabotropic and ionotropic effects \cite{Semyanov2001, Contractor2011}. Based on experiment facts, many computational models have been proposed taking into account neuron to astrocyte interactions to describe the interneuronal communication \cite{Nadkarni2004, Nadkarni2007, Volman2007, Perea2009, Gordleeva2012, Lazarevich2017, gordleeva2019astrocyte}. Many experimental works are shown that astrocytes can coordinate the neuronal network activations \cite{gordleeva2019astrocyte, Postnov2007, Wade2011, Amiri2012}. Because astrocyte is affected by a large number of synapses, the gliatransmission should also contribute to the effect of neuronal synchronization. Particularly, it was demonstrated in a hippocampal network, where calcium elevations in astrocytes and subsequent glutamate release led to the synchronous excitation of clusters of pyramidal neurons \cite{Angulo2004, Fellin2009}.

Coronavirus COVID-19 has become a global challenge of the modern world, stimulating intensive research in many related areas of science. Along with the development of vaccines, a fundamentally important global task is to investigate Covid-19 effects on different systems of human organisms. Recent studies have shown that coronavirus infection, entering the central nervous system and infecting astrocytes, causes various metabolic disorders, one of which is a decrease in the synthesis of astrocytic glutamate and GABA \cite{Crunfli2021}. It is also known that in the postcovid state, patients may suffer from symptoms of anxiety and impaired cognitive functions, which may be a consequence of disturbed brain rhythms. In this paper, we propose a mathematical model of impact SARS-CoV-2-infected astrocyte on the ability to synchronize neural networks and produce brain rhythms. We show that depending on the degree of disturbance in the synthesis of gliatransmitters neuronal network synchronization can be partially or completely suppressed.

\section*{Results}

First, let us consider how the astrocytes induced the appearance of quasi-synchronous bursting dynamics. If no astrocytic feedback is activated, e.g., $\gamma_{Y}=0$, the network showed asynchronous spontaneous firing due to uncorrelated noisy component of applied current, $I_{ext}$ stimulated all neurons (not shown in the figures). When the feedback is activated, $\gamma_{Y}>0$, the model starts to generate population burst discharges as illustrated in Fig. \ref{fig:3}. Similar to previous modeling studies \cite{gordleeva2019astrocyte, Postnov2007, Wade2011, Amiri2012} \textbf the astrocytes started to coordinate neuronal activity, inducing a certain level of coherence in the network firing. On the one hand, each astrocyte was activated integrating neuronal activity in its neighboring space. On the other hand, when astrocyte was activated it facilitated synchronously the activation of accompanying neurons within a certain area. In a result, neurons generated quasi-synchronous high-frequency burst discharges (Fig. \ref{fig:3}). These discharges were synchronized with peaks of extracellular glutamate concentration associated with the astrocytes activations. It should be noted that population burst dynamics is typical for living networks formed in dissociated cortical (or hippocampal) neuronal culture models {\em in vitro} \cite{Wagenaar2006,Stephens2012, Bisio2014}. In such biological models {\em normal} bursting indicates normal activity. In different pathological conditions (hypoxic–ischemic injury, alpha or theta coma or electrocerebral inactivity \cite{Johnson2019}) bursting fails what indicates the decrease of functional coherence in the network firing.

Next, we activated the virus pathological action in the model by increasing $\gamma_{virus}>0$. Figure \ref{fig:4} illustrates how network activity changed in this case. The raster plot shows that normal bursting were interrupted by the intervals of asynchronous uncorrelated firing. Corresponfing graphs of glutamate concentration in the right panels indicate that in these intervals the astrocytes were partly (lower peaks) or completely (no peaks) inhibited. After this intervals bursts were spontaneously recovered to normal sequences. So that, the result of SARS-CoV-2-infection at network level provokes to the failure of normal synchronization at network level while each neuron in the network works fine and each synaptic connections stay well functioning. Note, that for low values of $\gamma_{virus}$ associated with a ``light'' infection cases the intervals of uncorrelated firing are quite shot indicating a kind intermittent behavior between long lasting normal synchronous (e.g. ``laminar'') stages and rather shot pathological asynchronous (e.g. ``turbulent'') breaks.    

The next prediction of the model concerns a gradual character of the infection influence. It means that higher level of SARS-CoV-2 concentration in the organism will result in stronger pathological response. In terms of our model the increase of $\gamma_{virus}$ leads to the increase of intervals of ``pathological'' firing (Fig. \ref{fig:5}). One can note that the number of normal bursts withing the same sample window significantly descrease. In terms of neuro- and gliatransmitter concentrations (right panels of Fig. \ref{fig:5}) we also noticed the decrease of functionality not only of all astrocytes but also neurons. Some of them become depressed because of lack of sufficient amount glutamate to support normal excitatory transmission. So that, the higher SARS-CoV-2 concentration is exposed, then more astrocytes are infected and, hence, more ``explicit'' pathological synchrony breaks appear at the level of network firing.      

As one may expect now, further increase of $\gamma_{virus}$ completely inhibited the synchronization. It is illustrated in Fig. \ref{fig:6}. Correspondingly, all astrocytes failed to realease any glutamate. Note, however, that overall network firing 
still preserves sustained by activations of excitatory neurons with relatively strong glutamatergic synapses.    
To quantify the gradual character of  the network dysfunction due to SARS-CoV-2 infection we calculated the quantity reflecting the  average burst frequency versus $\gamma_{virus}$ (Fig. \ref{fig:7}). The graph represents monotonically descreading function vanishing at $\gamma_{virus} \rightarrow 1$.

\section*{Discussion}

We proposed a spiking neuron network model of synaptically coupled neurons accompanied of SARS-CoV-2-infected astrocytes. The model accounts for astrocyte activation depending on the integrative level of neuronal firing and the astrocyte to neuron feedback that is based on released gliatransmitter (glutamate) that facilitates group firing of neurons within the astrocyte territory. We found that the astrocyte disfunction and failure of gliatransmitter release that was the consequence of SARS-CoV-2-infection lead to failure of network synchronization. We have also illustrated that normal dynamics can be restored spontaneously, interspersed with intervals of pathological arousal.

At present, cognitive dysfunctions are reported as one of the most dangerous consequences of SARS-CoV-2 infection in post covid states. At the cognitive level, normal brain functioning can be associated with certain functional networks, where a particular function is associated with long-range correlations between different neuronal groups. Failure of such correlations may indicate the appearance of particular cognitive dysfunctions.

At the cellular level, functional synchronization is provided by coherent firing patterns of underlying spiking neuronal circuits. Following {\em in vitro} biological models of neuronal cultures where the appearance of population bursts provides functional synchronization, our mathematical model predicted that infected astrocytes might be responsible for failure of functional synchronization and consequent cognitive dysfunctions.

\section*{Methods}

\subsection*{Mathematical model of single neuron}
To describe the dynamics of a single neuron, we take Izhikevich's model \cite{Izhikevich2003}. It represents a compromise between computational complexity and biophysical plausibility. Despite its computational simplicity, this model can reproduce a large number of phenomena occurring in real neurons. The Izhikevich model is given in the form of a differential equations system (\ref{eq:01}):

\begin{equation}
\begin{cases}
C_{m}\frac{dV_{m}}{dt} = k(V_{m} - V_{r})(V_{m} - V_{t}) - U_{m} + I_{ext} + I_{syn} ,\\
\frac{dU_{m}}{dt} = a(b(V_{m} - V_{r}) - U_{m}).\label{eq:01}
\end{cases}
\end{equation}

If $V_{m} \geq V_{peak}$, than
\begin{equation}
\begin{cases}
V_{m} = c, \\
U_{m} = U_{m} + d, \label{eq:02}
\end{cases}
\end{equation}
where $a, b, c, d, k, C_{m}$ are the different parameters of the neuron. $V_{m}$ is the potential difference on the inside and outside of the membrane, and $U_{m}$ is a "recovery variable" describing the process of activation and deactivation of potassium and sodium membrane channels, respectively. As a result, we have negative feedback concerning the dynamics of the potential $V_{m}$ on the cell membrane. The resting potential value in the model lies in the range from –70 to –60 mV. Its value is determined by the parameter $b$, which describes the sensitivity of the recovery variable to subthreshold potential fluctuations on the neuronal cell membrane. The parameter $a$ sets the characteristic time scale of the change in the recovery variable $u$. The $V_{peak}$ value limits the spike amplitude. Parameters $c$ and $d$ specify the values of $V_{m}$ and $u$ after spike generation. $I_{ext}$ is the externally applied current. The neuron model is in an excitable mode and will demonstrate the generation of spikes in response to an applied current.  $I_{syn}$ is the sum of synaptic currents from all neurons with which this neuron is connected. Synaptic currents were calculated as follows:

\begin{equation}
I_{syn} = \sum y_{ij}w_{ij},  \label{eq:03}
\end{equation}
so that $I_{syn}$ represents the weighted sum of all synaptic currents of postsynaptic neurons with $w_{ij}$ denoting the weights for glutamatergic and GABAergic synapses between neurons. For excitatory and inhibitory contacts, the weights have positive and negative signs, respectively. Variables $y_{ij}$ denote the output signal (synaptic neurotransmitter) from the $i$th neuron to the $j$th neuron which involved in generation of $I_{syn}$.
In our model, the number of synaptic connections is $N^{2} \times p$, where $N$ is the number of neurons, $p$ is the probability of communication between two random neurons and equal 0.1 ($10 \%$ of connections). Each synaptic weight was set randomly for all connections in the range from 20 to 60. If a spike is generated on the presynaptic neuron, a jump in the synaptic current occurs on the postsynaptic one, which further decays exponentially. As a result, synaptic neurotransmitter concentration, $y_{ij}$, was calculated as follows:
\begin{equation}
y_{ij}(t) = 
\left\{
\begin{array}{lcl}
y_{ij}(t_i)exp(-t/\tau_y)  & \mbox{if}, & t_s<t<t_{s+1},  \\
    y_{ij}(t_s-0)+1 & \mbox{if}, & t=t_s,
\end{array}
 \right.
  \label{eq:0a}
\end{equation}
where $t_s$ denotes the time moments of consequent presynaptic spikes, $\tau_y$ is a relaxation time constant.

Each spike in the neuron model induces the release of neurotransmitter. To describe the neuron to astrocyte cross-talk, here we only focus on the excitatory neurons releasing glutamate. Following earlier experimental and modeling studies, we assumed that the glutamate-mediate exchange was the key mechanism to induce coherent neuronal excitations \cite{Angulo2004, Fellin2009}. The role of GABAergic neurons in our network is to support the excitation and inhibition balance avoiding hyperexcitation states.

For simplicity, we take a phenomenological model of released glutamate dynamics. In the mean field approximation average concentration of extyrasynaptic glutamate concentration for each excitatory synapses, $X_{e}$, was described by this equations:
\begin{equation}
X_{e}(t) = 
\left\{
\begin{array}{lcl}
X_{e}(t_s)exp(-t/\tau_X),  & \mbox{if} & t_s<t<t_{s+1},  \\
    X_{e}(t_s-0)+1, & \mbox{if} & t=t_s,
\end{array}
 \right.
  \label{eq:0b}
\end{equation}
where $e=1,2,3,...$ is the index of excitatory presynaptic neurons, $s=1,2,3,\ldots$ is the index of the presynaptic spikes, $\tau_{X}$ is the time relaxation. After the spike is generated on the presynaptic neuron, the neurotransmitter is released, and the concentration of the extrasynaptic neurotransmitter increases due to diffusion processes, which decreases over time with its characteristic time,$\tau_{X}$. So that, the difference in mathematical descriptions of synaptic (\ref{eq:0a}) and extrasynaptic (\ref{eq:0b}) was accounted by different time constants $\tau_y$ and $\tau_X$, respectively.

\subsection*{Astrocytic dynamics}

Part of the extrasynaptic glutamate can bind to metabotropic glutamate receptors of the astrocyte processes. Next, after a cascade of molecular transformations mediated by elevation of intracellular calcium the astrocyte release of gliatransmitter back to the extracellular space. For our purpose, in mathematical model we dropped detailed description of these transformations, defining only input-output functional relation between the neurotransmitter and gliatransmitter concentrations in the following form \cite{stasenko2020quasi,Gordleeva2012}: 
\begin{equation}
\frac{dY_{e}}{dt} = - \alpha_{Y}Y_{e} + \frac{\beta_{Y} (1 - \gamma_{virus})}{1+exp(-X_{e} + X_{thr})} \label{eq:04}
\end{equation}
where $e=1,2,3,\ldots$ is the index of excitatory neuron,  $Y_e$ is the gliatransmitter concentration in the neighborhood of corresponding excitatory synapse, $\alpha_{Y}$ is the clearance rate. 
The second term in Eq. (\ref{eq:02}) describes the gliatransmitter production when the mean field concentration of gliatransmitter exceeds some threshold, $X_{thr}$. Figure 1 illustrates the network construction and neuron to astrocyte crosstalk for excitatory glutamatergic synapses. 

Based on experimental facts demonstrated that Covid-19 infection resulted in the decrease of astrocytic glutamate and GABA synthesis \cite{Crunfli2021} we accounted it by the coefficient of infection of astrocytes $0< \gamma_{virus}<1$. For a well-functioning cell, it takes unity value and the production rate is accounted by $\beta_Y$, while for the dead cell, it takes zero value.

\subsection*{Astrocytic modulation of neural activity}

It follows from experimental facts that astrocytes can influence on probability of neurotransmitter release \cite{Perea2007, Jourdain2007,Fiacco2004}. 
In turn, it results in modulation synaptic currents. We accounted this in the following form for glutamatergic synapses: 
\begin{equation}
I_{syn} = \sum y_{ij}w_{ij} (1+\gamma_{Y} \cdot Y_e) \label{eq:05}
\end{equation}
where $I_{syn}$ is the summation of all synaptic currents of postsynaptic neuron, $w$ is the weight for glutamatergic synapses between neurons, $\gamma_{Y}$ is the coefficient of astrocyte influence on synaptic connection. 

\subsection*{Neural Network}

\begin{figure}[h!]
\begin{center}
\includegraphics[width=18cm]{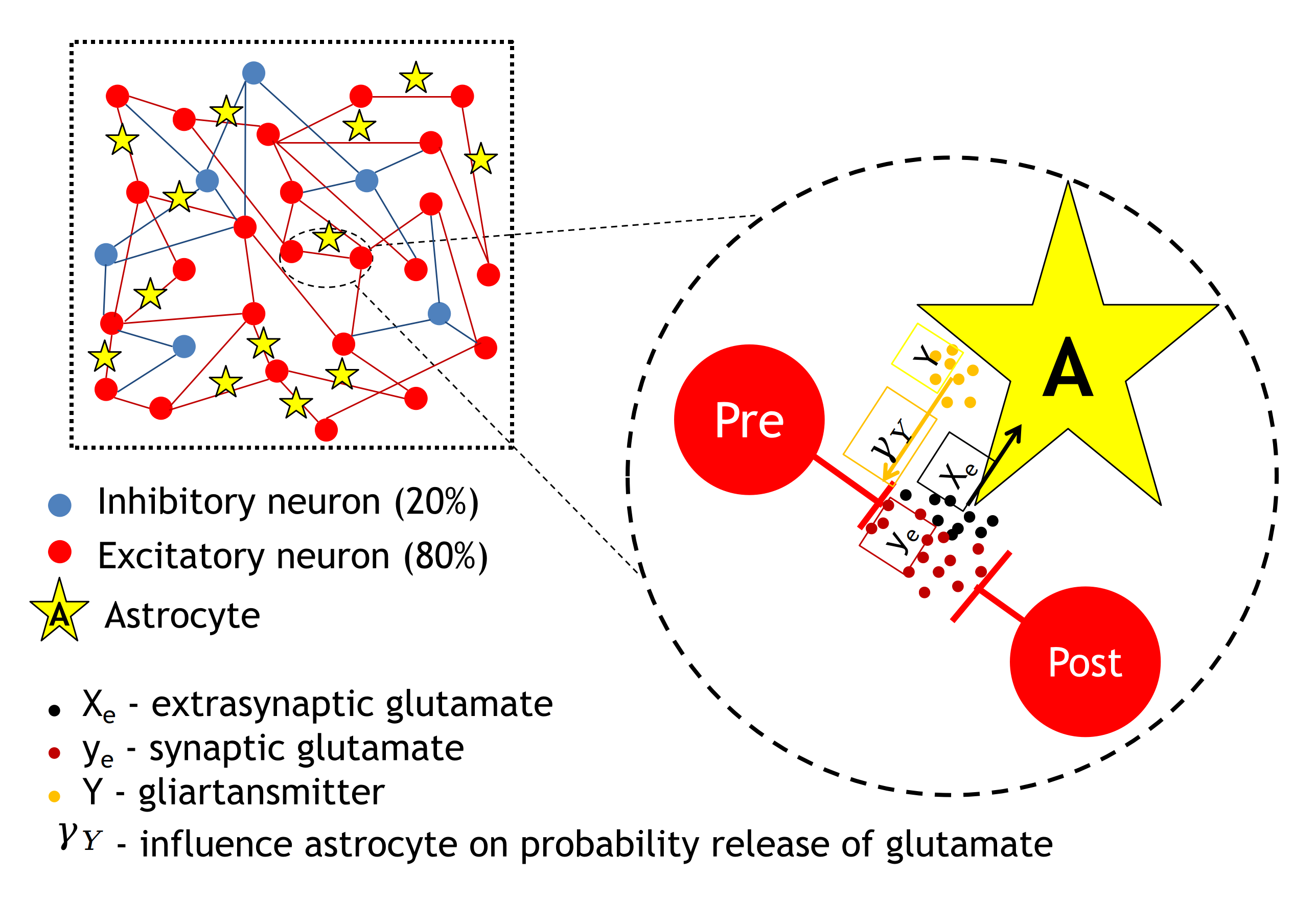}
\end{center}
\caption{ Schematic view of the network and schematic representation of astrocytic modulation of synaptic current.}\label{fig:1}
\end{figure}

Schematic representation of network with astrocytic modulation of probability release of neurotransmitter is presented in Fig.\,\ref{fig:1}. After the generation of an action potential on the presynaptic neuron, neurotransmitter is released from the presynaptic terminal. Its part can diffuse out of the cleft where it can bind to specific astrocyte receptors \cite{Rusakov1998}. The activation of the astrocyte results in the generation of calcium transients in the form of short-term increase in the intra- cellular concentration of calcium. In turn, the calcium elevations lead to gliotransmitter (particularly glutamate) release. The released gliatransmitter, reaching the presynaptic terminal, leads to a change in the probability of neurotransmitter release, potentiating the synaptic current. This, in turn, leads to the formation of burst activity. 

\begin{figure}[h!]
\begin{center}
\includegraphics[width=18cm]{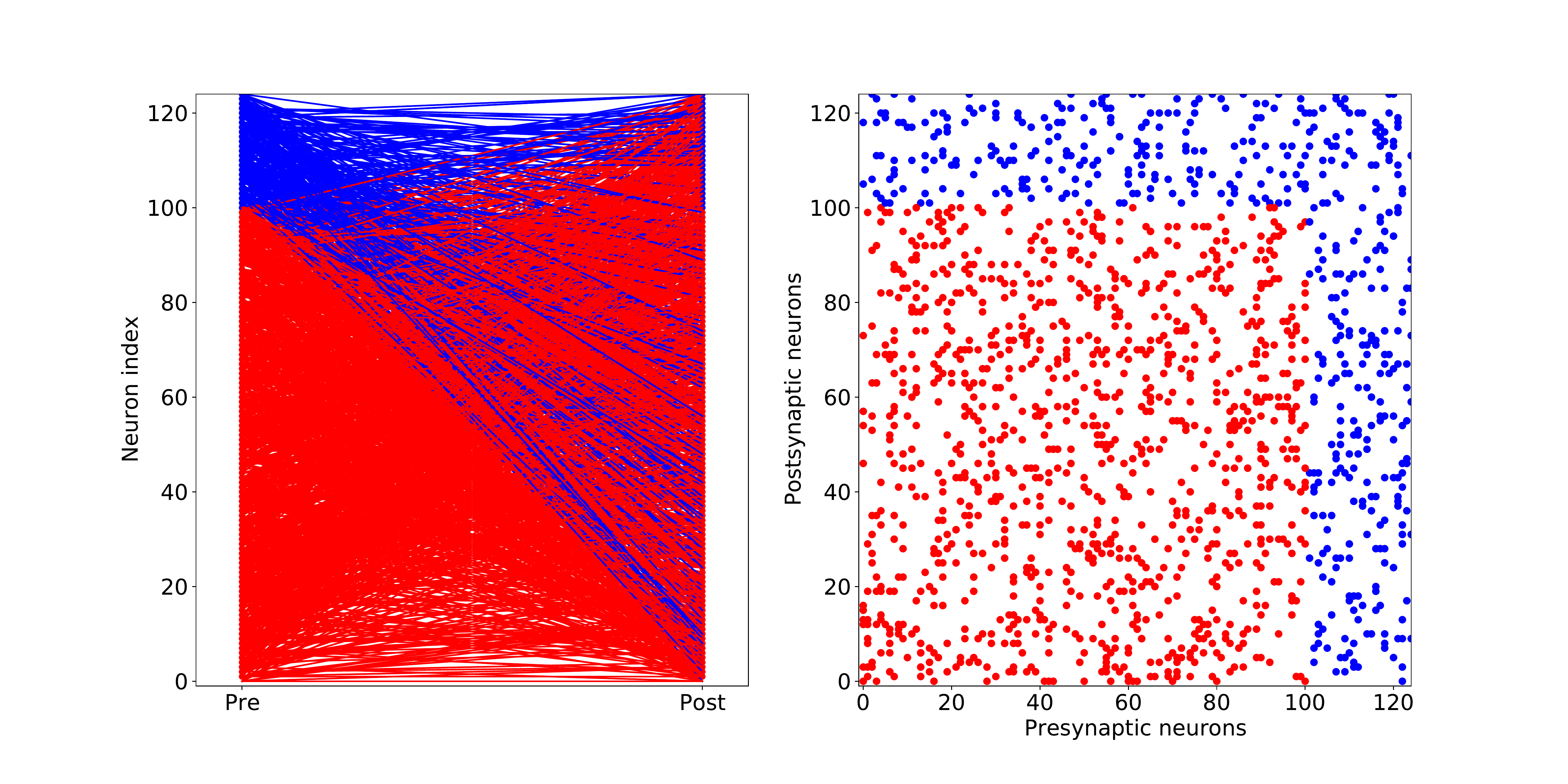}
\end{center}
\caption{Scheme of synapse connections in neural networks. The left panel shows the connections between pre and postsynaptic neurons. Neurons on the vertical axis are ordered with excitatory ones, $N_{ex}$, coming first followed by the inhibitory ones, $N_{inh}$. The synaptic connections are illustrated by lines from the left (``Pre'')  to the right (``Post'') in the figure.  Red lines denote the excitatory connections, the blue lines correspond to the inhibitory ones. The figure on the right shows connectivity matrix, $w_{ij}$, with coordinates according to the numbers of pre- and postsynaptic neurons. Each dot in the field denotes the presence of nonzero synaptic connections.}\label{fig:2}
\end{figure}

\begin{figure}[h!]
\begin{center}
\includegraphics[width=18cm]{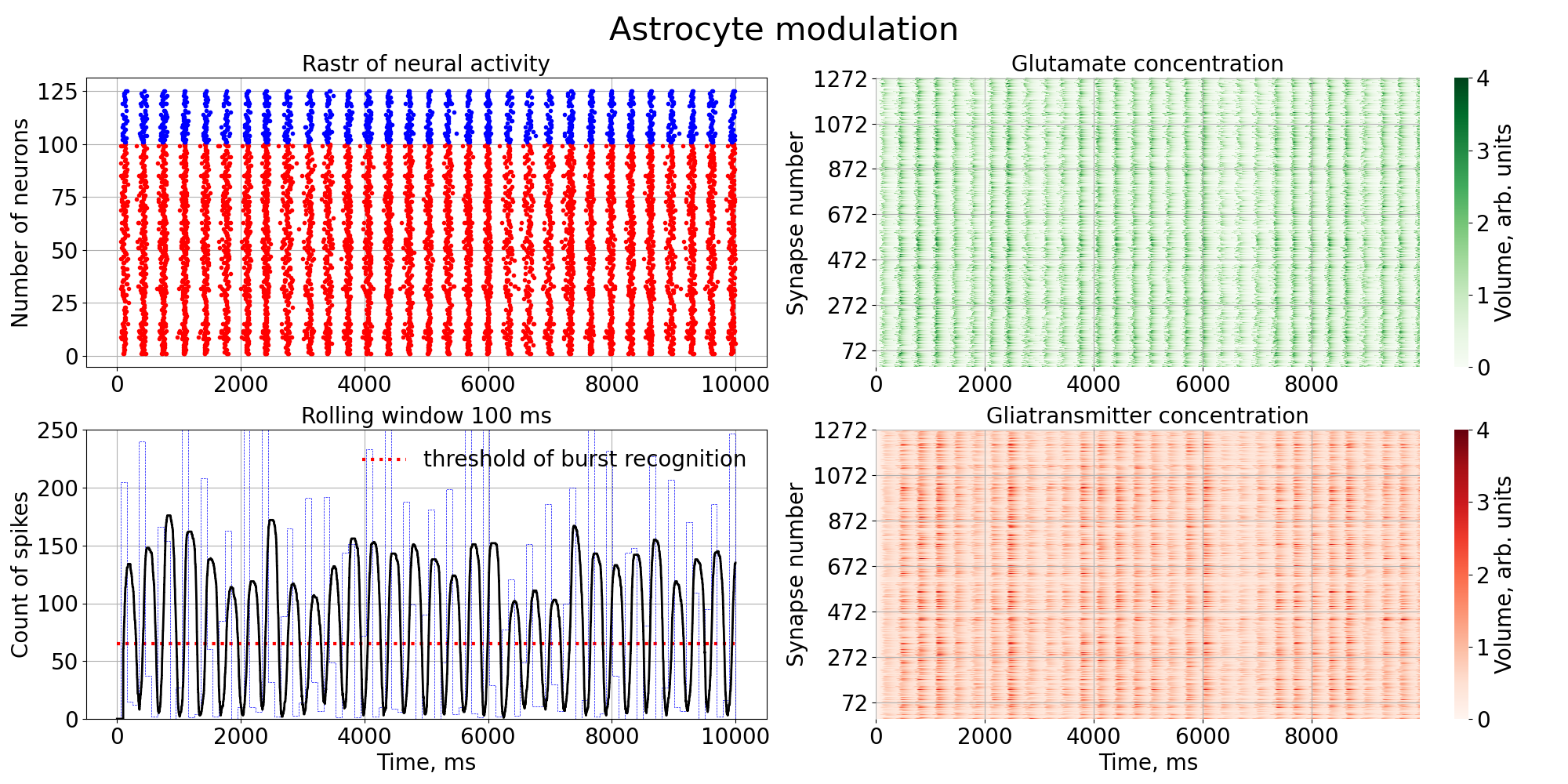}
\end{center}
\caption{Network firing in normal conditions. Left upper panel: Raster plot  of neural activity. The red dots show spikes by excitatory (glutamatergic) piramidal neurons and the blue onces by inhibitory (GABAergic) interneurons. 
Right panels: Changes in the extracellular concentrations of synaptic glutamate diffused from the cleft  (green color) and the glutamate released by astrocyte (red color) for all tripartite synapse. Left lower panel: Average spiking rate over a sliding time window of 100 ms from the entire simulation time of the model. We set the burst generation threshold at 65 spikes marked by a red dashed line. Parameter values: for neuron -  $a =  0.02, b = 0.5, c = -40, d = 100, k = 0.5, C_{m} = 50, V_{r} = -60, V_{peak} = 35, V_{0} = -60, U_{0} = 50 $;  other - $\tau_y = 4, \tau_X = 100, \alpha_{Y} = 80, \beta_{Y} = 1, X_{thr} = 5.6, \gamma_{Y} = 0.72 $.}\label{fig:3}
\end{figure}

\begin{figure}[h!]
\begin{center}
\includegraphics[width=18cm]{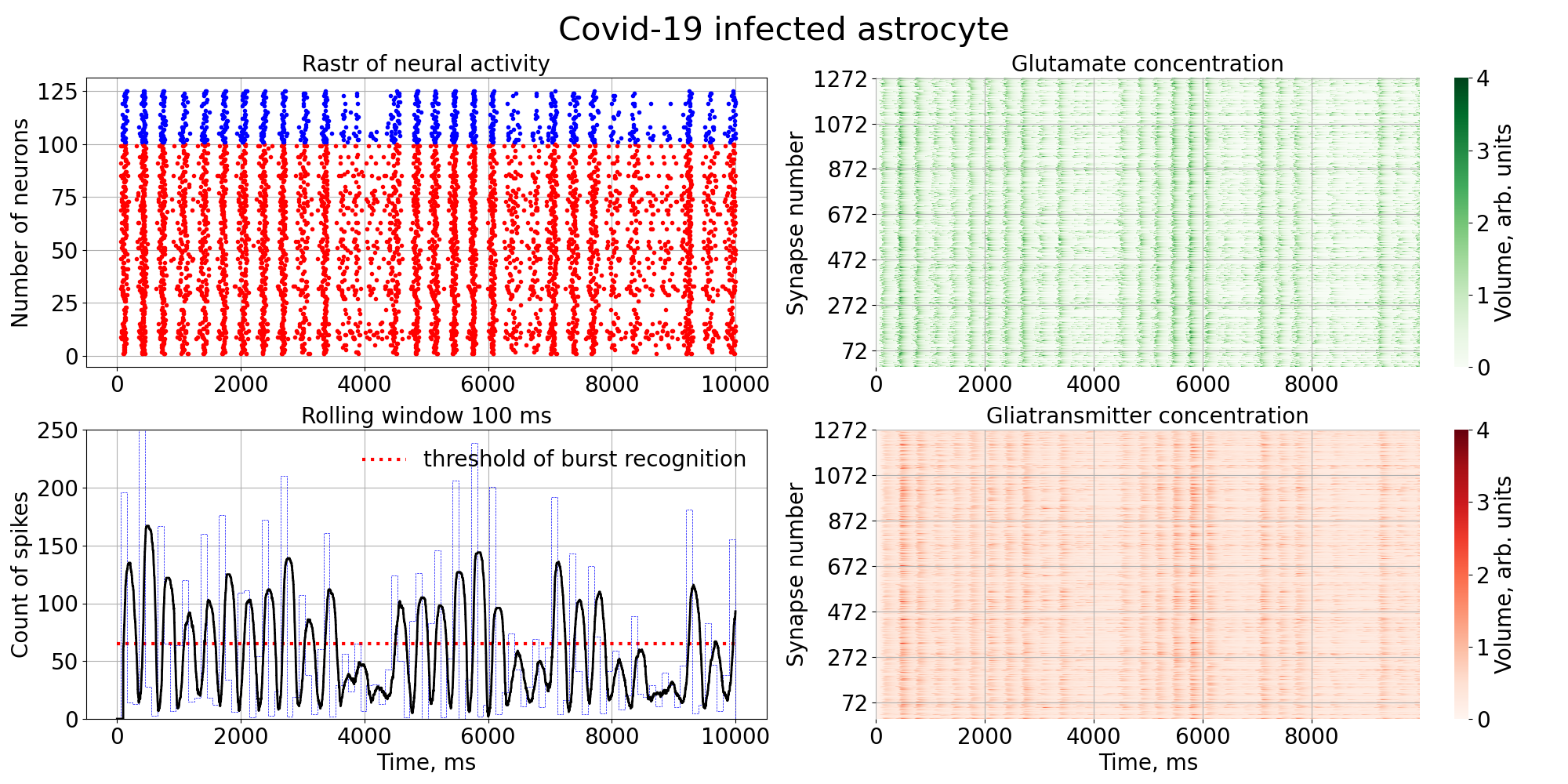}
\end{center}
\caption{Raster chart  of neural activity  and dependences of  gliatransmitter  and neurotransmitter concentration from time  with Covid-19 infected astrocyte feedbacks for $\gamma_{virus}=0.10$}\label{fig:4}
\end{figure}

\begin{figure}[h!]
\begin{center}
\includegraphics[width=18cm]{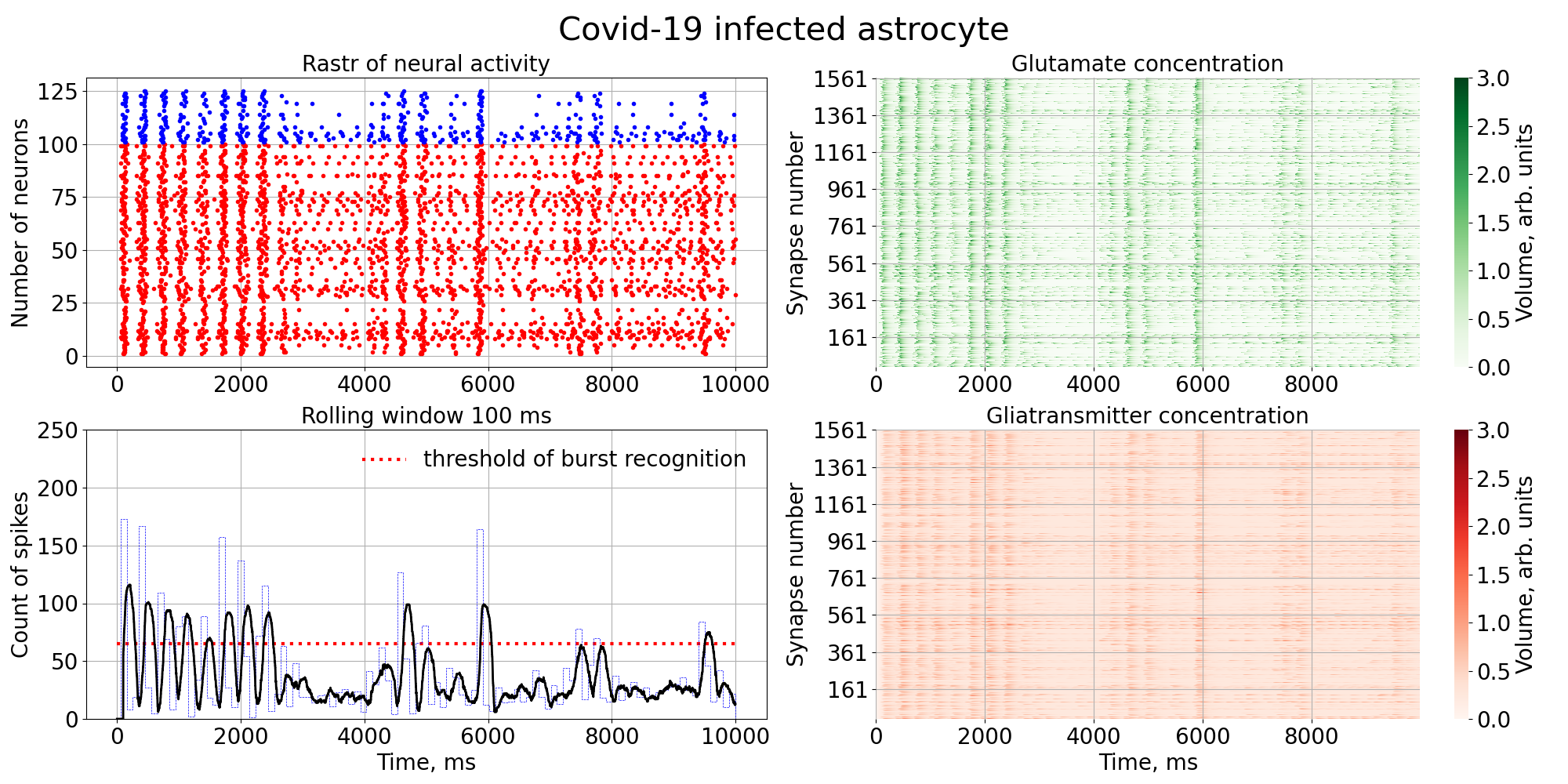}
\end{center}
\caption{Raster chart  of neural activity  and dependences of  gliatransmitter  and neurotransmitter concentration from time  with Covid-19 infected astrocyte feedbacks for $\gamma_{virus}=0.2$}\label{fig:5}
\end{figure}

\begin{figure}[h!]
\begin{center}
\includegraphics[width=18cm]{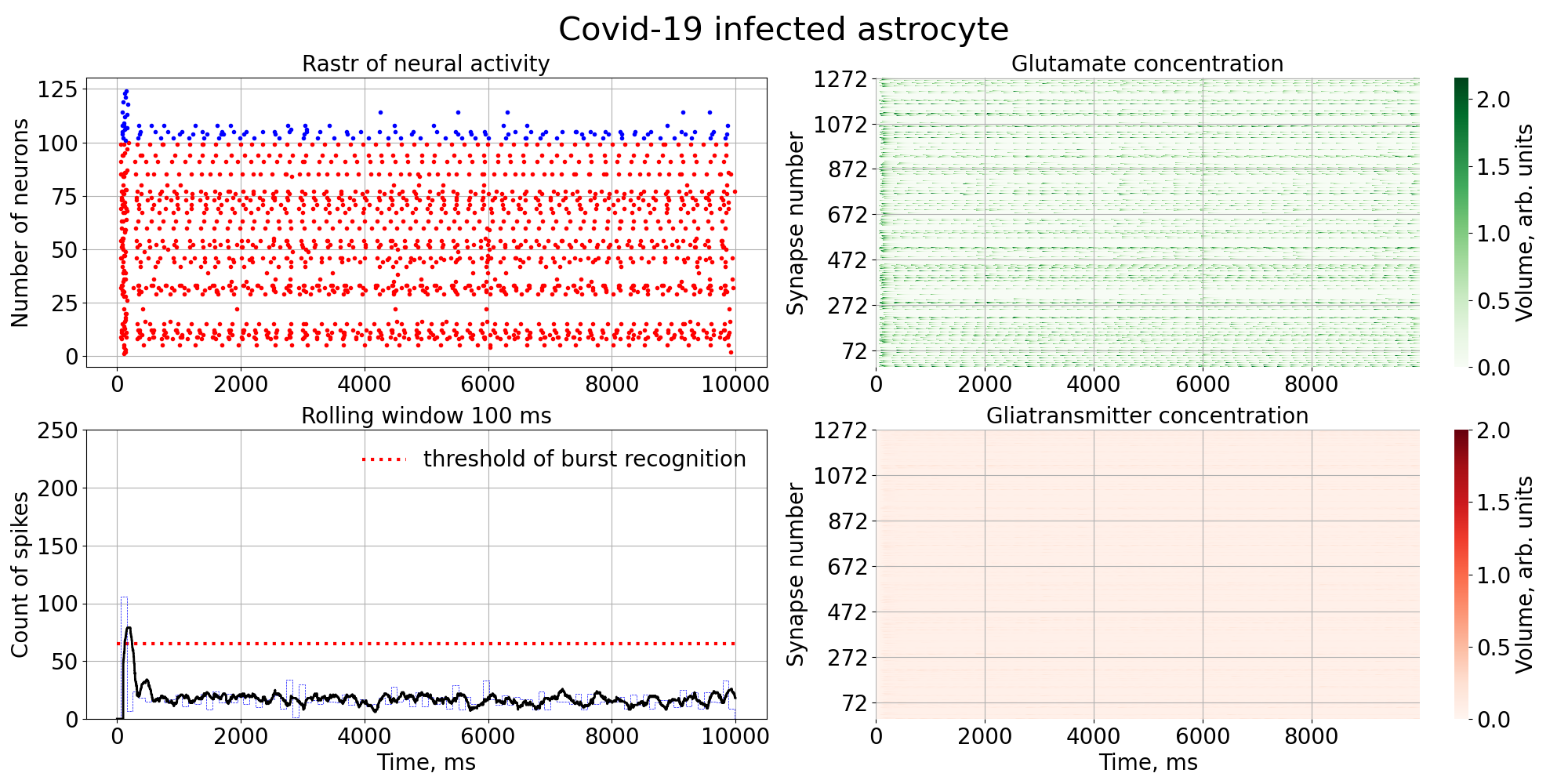}
\end{center}
\caption{Raster chart  of neural activity  and dependences of  gliatransmitter  and neurotransmitter concentration from time  with Covid-19 infected astrocyte feedbacks for $\gamma_{virus}=0.8$}\label{fig:6}
\end{figure}

\begin{figure}[h!]
\begin{center}
\includegraphics[width=18cm]{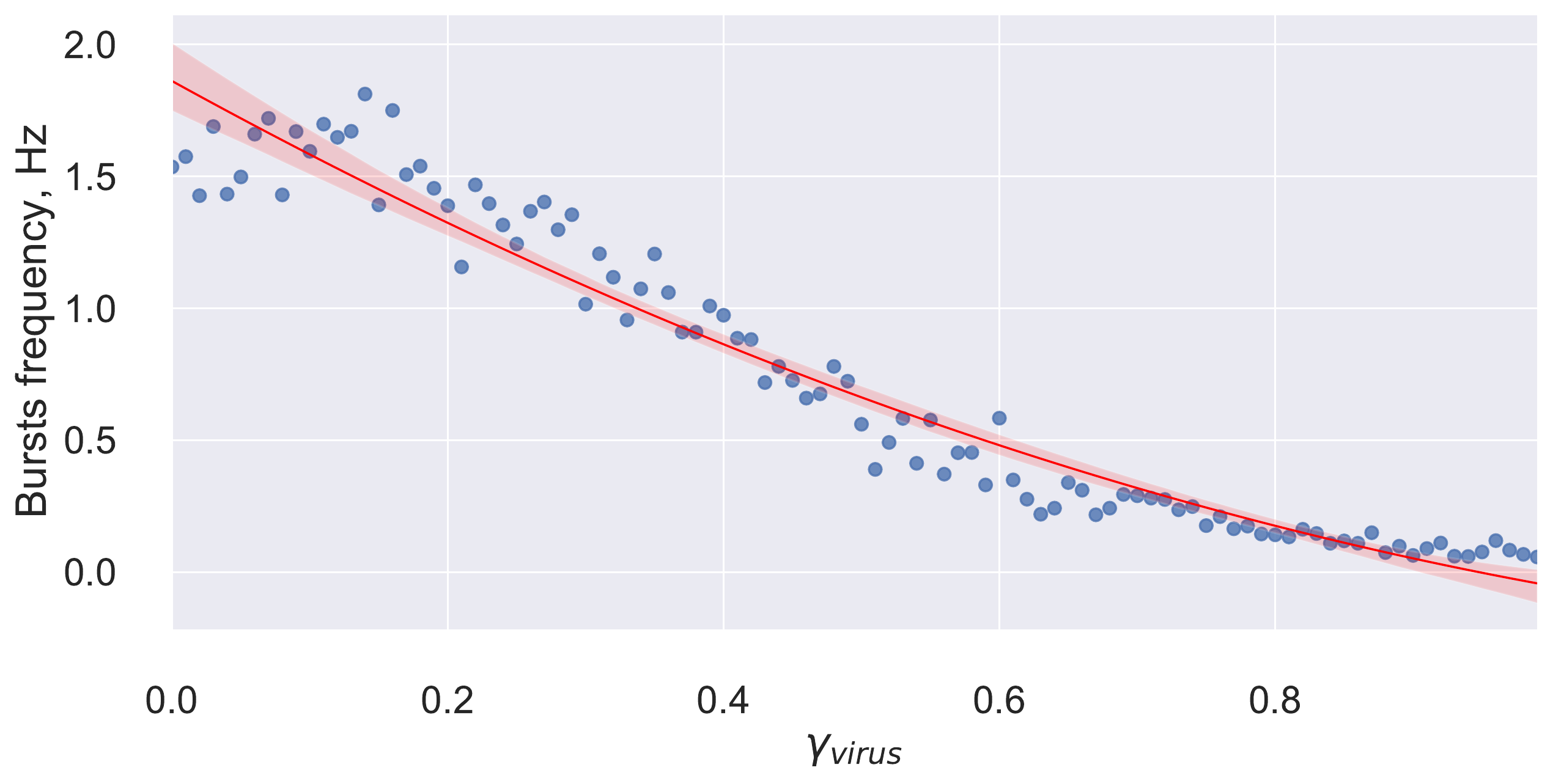}
\end{center}
\caption{Dependence of bursts frequency from $\gamma_{virus}$.}\label{fig:7}
\end{figure}

The architechture of synaptic connections in our model neywork is illustrated in figure \ref{fig:2}. The left panel shows the connections between pre and postsynaptic neurons. Neurons on the vertical axis are ordered with excitatory ones, $N_{ex}$, coming first followed by the inhibitory ones, $N_{inh}$. The synaptic connections are illustrated by lines from the left (``Pre'')  to the right (``Post'') in the figure.  Red lines denote the excitatory connections, the blue lines correspond to the inhibitory ones. The figure on the right shows connectivity matrix, $w_{ij}$, with coordinates according to the numbers of pre- and postsynaptic neurons. Each dot in the field denotes the presence of nonzero synaptic connections.

In our simulations we used $N=125$ spiking cortical neurons with 1562 synaptic connections in real time (resolution 1 ms). Motivated by the anatomy of a mammalian cortex, we choose the ratio of excitatory to inhibitory neurons to be 4 to 1. So that, we take $N_{ex}=100$ and $N_{inh}=25$, respectively. Besides the synaptic input, each neuron receives a noisy thalamic input ($I_{ext}$). The noisy thalamic input is set in a random way for all neurons in the range from 0 to 50. Since the model uses a mean-field approach to describe changes in the main neuroactive substances (neurotransmitter and gliatransmitter), we do not separate the effect of a single astrocyte on a group of neurons or a group of neurons on a single astrocyte, but we introduce into the description of each synaptic contact its own dynamics for the neuro and gliatransmitter.

We performed the numerical integration of the model (1)–(7) using the Euler method with a step of 0.5 ms. Such a procedure has been shown to be appropriate for integrating large systems of the Izhikevich’s neurons \cite{Izhikevich2003, izhikevich2004model}. To simulate the model, software was written in the object-oriented programming language C++.

\bibliography{biblio}

\section*{Acknowledgements}

This work was partially funded by the Russian Ministry of Science and Education project \# №075-15-2021-634 (data analysis and numerical simulation of the model) and Development Programs of the Regional Scientific and Educational Mathematical Center "Mathematics of Future Technologies" project №075-02-2020-1483/1 (development of mathematical model).

\section*{Author contributions statement}

S.V.S designed the research and idea. S.V.S. simulated the model. S.V.S. performed data
analysis. S.V.S., V.B.K., and A.E.H. interpreted the results. S.V.S. and V.B.K. formulated the model. All authors participated in writing and editing the manuscript.

\section*{Competing interests}
The authors declare no competing interests.

\section*{Code availability}
Code used to produce the results presented herein is available in a public GitHub repository at https://github.com/sstasenko/

\section*{Additional information}

Correspondence and requests for materials should be addressed to S.V.S.

\end{document}